\documentclass[spanish,12pt]{article}
\usepackage{amsmath}
\usepackage{amsfonts}
\usepackage{latexsym}
\usepackage{graphicx}
\usepackage{multicol}
\usepackage{color}
\usepackage[usenames,dvipsnames,svgnames,table]{xcolor}
\usepackage{url}
\usepackage[english]{babel}
\usepackage[utf8]{inputenc}

\input tcilatex

\begin{document}

\title{The duality problem in symbolic principal component for interval-valued variables}

\author{ Oldemar Rodr\'{\i}guez \thanks{University of Costa Rica, San Jos\'e, Costa Rica;
E-Mail: oldemar.rodriguez@ucr.ac.cr}}
\date{Recibido: 20 de marzo 2015}

\maketitle

\begin{abstract}
In \cite{ref11}  and \cite{ref3}, the authors proposed the Centers and the Vertices Methods to extend the well known principal components analysis method to a particular kind of symbolic objects
characterized by multi--valued variables of interval type. Nevertheless the authors
use the classical circle of correlation to represent the variables. The
correlation between the variables and the principal components are not
symbolic, because they compute the standard correlations between the midpoints of variables and the midpoints 
of the principal components.

It is well known that in standard principal component analysis we may compute the
correlation between the variables and the principal components using the
duality relations starting from the coordinates of the individuals in the
principal plane, also we can compute the coordinates of the individuals in
the principal plane using duality relations starting from the correlation
between the variables and the principal components. 

In this paper we propose
a new method to compute the symbolic correlation circle using duality
relations in the case of interval-valued variables. Besides, the reader may use all the methods presented herein and verify the results using the {\tt RSDA} package written in {\tt R} language, that can be downloaded and installed directly from {\tt CRAN}  \cite{Rod2014},
\end{abstract}

Symbolic data analysis,  interval principal components analysis, correlations circle, duality
relations .
\medskip


\section{Introduction}

Statistical and data mining methods have been developed mainly in the case in which variables take a single value. Nevertheless, in real life there are many situations in which the use of this type of variables may cause an important loss of information or reduction in quality and veracity of results. In the case of quantitative variables, a more complete information can be achieved by describing an ensemble of statistical units in terms of interval data, that is, when the value taken by a variable is an interval  $[a,b]$ with $a,b \in \mathbb{R}$. 

An especially useful case where it is convenient to summarize large ensembles of data in such a way that the summary of data resulting is of a more manageable size, which in turn maintains the greatest amount of information it had in the original data set. In this problem, the central idea is to substitute the ensemble of all transactions carried out by a person or client (for example the owner of a credit card) for one only “transaction” that summarizes all originals in such a way that millions of transactions could be summarized in an only one that maintains the client’s habitual behavior. The above is achieved thanks to this new transaction will have in its fields not only numbers (as in the usual transactions), but will also have intervals that store, for example, the minimum and maximum purchase. In experimental evaluation section, we will provide an example that illustrates these ideas for which we will use the ``US Communities and Crime Data Set''  \cite{Bache2013}.

The statistical treatment of the interval-type data has been considered in the context of Symbolic Data Analysis – SDA) introduced by E. Diday in \cite{Diday1987}, the objective of which is to extend the classic statistical methods to the study of more complex data structures that include, among others, interval-valued variables. A complete presentation on Symbolic Data Analysis can be found in the following works \cite{BockDiday2000,BillardDiday2003,BillardDiday2006}.

As it is very well know, Principal Components Analysis (PCA) is a way of identifying patterns in data, and expressing the data in such a way as to highlight their similarities and differences. Since patterns in data can be hard to find in data of
high dimension PCA is a powerful tool for analyzing data. In Data Mining, the other main advantage of PCA is that once you have found these patterns in the data, and you compress the data, i.e. by reducing the number of dimensions, without
much loss of information.  In the framework of Symbolic Data Analysis, different authors have proposed some approaches oriented to extend principal component analysis method  to study the relationships between symbolic objects. The first approaches, Vertices PCA and Centers PCA, were proposed by Chouakria in \cite{ref4} and by Cazes et al. in \cite{ref3}. A second approach was proposed by Lauro and Palumbo in \cite{Lauro2000}, they emphasize the fact that a box is a cohesive set of vertices that depends also on its size and shape, introducing a more consistent way to treat units as a complex data representation by introducing suitable constraints for the vertices belonging to the same object. Other approaches to principal component analysis for interval-valued variables have been proposed in \cite{Lauro2005}, \cite{Lauro2006}, \cite{Gioia2006}.
\section{The duality problem in Centers Method}

\markright{The duality problem in Centers Method}

In Principal Components Analysis to interval variables the input is $m$
symbolic objects $S_1,S_2,\ldots ,S_m$ describe by $n$ interval variables $%
X^1,X^2,\ldots ,X^n$ like we show in the equation (1). 

\begin{equation}
\left(
\begin{array}{c}
S_1 \\
\vdots  \\
S_m
\end{array}
\right) =\left(
\begin{array}{ccc}
X_{S_11} & \cdots  & X_{S_1n} \\
\vdots  & \ddots  & \vdots  \\
X_{S_m1} & \cdots  & X_{S_mn}
\end{array}
\right) =\left(
\begin{array}{ccc}
\left[ \underline{x_{11}},\overline{x_{11}}\right]  & \cdots  & \left[
\underline{x_{1n}},\overline{x_{1n}}\right]  \\
\vdots  & \ddots  & \vdots  \\
\left[ \underline{x_{m1}},\overline{x_{m1}}\right]  & \cdots  & \left[
\underline{x_{mn}},\overline{x_{mn}}\right]
\end{array}
\right) ,  \label{eq:eqcc41}
\end{equation}

The idea of the centers method is to transform the matrix presented in (1) in the following matrix (2):

\begin{equation}
X^c=\left(
\begin{array}{cccc}
x_{11}^c & x_{12}^c & \cdots & x_{1n}^c \\
x_{21}^c & x_{22}^c & \cdots & x_{2n}^c \\
\vdots & \vdots & \ddots & \vdots \\
x_{m1}^c & x_{m2}^c & \cdots & x_{mn}^c
\end{array}
\right) =\left(
\begin{array}{cccc}
\frac{\underline{x_{11}}+\overline{x_{11}}}2 & \frac{\underline{x_{12}}+%
\overline{x_{12}}}2 & \cdots & \frac{\underline{x_{1n}}+\overline{x_{1n}}}2
\\
\frac{\underline{x_{21}}+\overline{x_{21}}}2 & \frac{\underline{x_{22}}+%
\overline{x_{22}}}2 & \cdots & \frac{\underline{x_{2n}}+\overline{x_{2n}}}2
\\
\vdots & \vdots & \ddots & \vdots \\
\frac{\underline{x_{m1}}+\overline{x_{m1}}}2 & \frac{\underline{x_{m2}}+%
\overline{x_{m2}}}2 & \cdots & \frac{\underline{x_{mn}}+\overline{x_{mn}}}2
\end{array}
\right) ,  \label{eq:eqd1}
\end{equation}

\noindent then in the centers method we apply the standard principal components
analysis to the matrix (2). To apply this standard principal
components in \cite{ref3} the authors use the matrix of
variance--covariance $V^c=(X^c)^tX^c$ and then to compute the interval
principal component $[\underline{y_{ik}},\overline{y_{ik}}]$
they have proposed the equations (3) and (4).

\begin{equation}
\underline{y_{ik}}=\dsum\limits_{j,u_{jk<0}}\left( \overline{x_{ij}}-%
\overline{X_j^c}\right) u_{jk}+\dsum\limits_{j,u_{jk>0}}\left( \underline{%
x_{ij}}-\overline{X_j^c}\right) u_{jk},  \label{eq:eqd2}
\end{equation}

\begin{equation}
\overline{y_{ik}}=\dsum\limits_{j,u_{jk<0}}\left( \underline{x_{ij}}-%
\overline{X_j^c}\right) u_{jk}+\dsum\limits_{j,u_{jk>0}}\left( \overline{%
x_{ij}}-\overline{X_j^c}\right) u_{jk}.  \label{eq:eqd3}
\end{equation}

\noindent where $\overline{X_j^c}$ is the mean of the column $j$--th of the matrix $%
X^c $, and $u=(u_{1k},u_{2k},\ldots ,u_{nk})$ is the $k-$th eigenvector of $%
V^c$.

We are going to center and reduce the matrix $X^c$ in order to work with
correlations as we show in (5) where $\overline{X_j^c}$ and $%
\sigma _j^c$ are the mean and the variance of the column $j$--th of the
matrix $X^c$ respectively:

\begin{equation}
z_{ij}=\frac 1{\sqrt{m}}\frac{x_{ij}^c-\overline{X_j^c}}{\sigma _j^c}.
\label{eq:eqd4}
\end{equation}

Then we will work with the matrix $Z=(z_{ij})_{\QATOPD. . {i=1,2,\ldots
,m}{j=1,2,\ldots ,n}}$. If we denote $z^j$ the column $j$--th of the matrix $%
Z$, so we have $(z^j)^t\cdot z^i=R(z^j,z^i)\leq 1$ then the center of the
hypercube variable is always inside of the radius one circle. We denote by $\overline{z}_{ij}^c=\frac 1{\sqrt{m%
}}\frac{\overline{x_{ij}^{}}-\overline{X_j^c}}{\sigma _j^c}$ and $\underline{%
z}_{ij}^c=\frac 1{\sqrt{m}}\frac{\underline{x_{ij}^{}}-\overline{X_j^c}}{%
\sigma _j^c}$.

The inertia matrix $ZZ^t$ is symmetrical, its eigenvectors are orthonormals
and its eigenvalues all are positive. We denote by $v_1,v_2,\ldots ,v_q$ the
$q$ eigenvectors of $ZZ^t$ associated to the eigenvalues $\lambda _1\geq
\lambda _2\geq \cdots \geq \lambda _q>0$. We also denote by $%
V=[v_1|v_2|\cdots |v_q]$ the matrix of size $m\times q$ that has as a
columns the eigenvectors of $ZZ^t$. It is well known that we can compute the
coordinates of the variables in circle of correlation by $Z^tV$, then we can
compute the coordinate of the $i$--th column of $X^c$ (point
centre--variable) on $j$--th component principal (on the direction of $v_j$)
by the equation (6):

\begin{equation}
r_{ij}=\dsum\limits_{k=1}^mz_{ki}v_{kj}.  \label{eq:eqd5}
\end{equation}

Like $Z$ is the matrix $X$ centered and reduced, the number $r_{ij}$ also
represents the correlation between the center of gravity of the
interval--variable $X^i$ and the $j$--th component principal.

\begin{theorem}
If we project the hypercube variable defined by the $i$--th column of $Z$ on
the $j$--th component principal (on the direction of $v_i$), then we have
that the minimum and the maximum value are given by the equation (7) and (8) respectively:

\begin{equation}
\underline{r_{ij}}=\dsum\limits_{k=1,v_{kj}<0}^m\overline{z}%
_{ki}^cv_{kj}+\dsum\limits_{k=1,v_{kj}>0}^m\underline{z}_{ki}^cv_{kj},
\label{eq:eqd6}
\end{equation}

\begin{equation}
\overline{r_{ij}}=\dsum\limits_{k=1,v_{kj}<0}^m\underline{z}%
_{ki}^cv_{kj}+\dsum\limits_{k=1,v_{kj}>0}^m\overline{z}_{ki}^cv_{kj}.
\label{eq:eqd7}
\end{equation}
\end{theorem}

To prove that, let be $\widehat{z}_j=(\widehat{z}_{1j},\widehat{z}%
_{2j},\ldots ,\widehat{z}_{mj})\in Z_H^j$ (the hypercube defined by the $j$%
-th column of $Z$) then $\widehat{z}_{ij}\in [\underline{z}_{ij}^c,\overline{%
z}_{ij}^c]$ for all $i=1,2,\ldots ,m$ and $j=1,2,\ldots ,q$. We denote by $p%
\widehat{z}_{ij}$ the projection of $\widehat{z}_j$ on the axis factorial
with direction $v_i$.

Since $\widehat{z}_{ij}\in [\underline{z}_{ij}^c,\overline{z}_{ij}^c]$ we
have (9) and (10):

\begin{equation}
\underline{z}_{ki}^cv_{kj}\leq \widehat{z}_{ki}v_{kj}\leq \overline{z}%
_{ki}^cv_{kj}\text{ if }v_{kj}\geq 0,  \label{eq:eqd71}
\end{equation}

\begin{equation}
\underline{z}_{ki}^cv_{kj}\geq \widehat{z}_{ki}v_{kj}\geq \overline{z}%
_{ki}^cv_{kj}\text{ if }v_{kj}\leq 0.  \label{eq:eqd72}
\end{equation}

By definition $p\widehat{z}_{ij}=\dsum\limits_{k=1}^m\widehat{z}_{ki}v_{kj}$
then:
\[
p\widehat{z}_{ij}=\dsum\limits_{k=1}^m\widehat{z}_{ki}v_{kj}=\dsum%
\limits_{k=1,v_{kj>0}}^m\widehat{z}_{ki}v_{kj}+\dsum\limits_{k=1,v_{kj<0}}^m%
\widehat{z}_{ki}v_{kj}.
\]

So, using (9) and (10) we get:

\[
p\widehat{z}_{ij}\leq \dsum\limits_{k=1,v_{kj}<0}^m\underline{z}%
_{ki}^cv_{kj}+\dsum\limits_{k=1,v_{kj}>0}^m\overline{z}_{ki}^cv_{kj}=%
\overline{r_{ij}},
\]

and analogously we have:

\[
p\widehat{z}_{ij}\geq \dsum\limits_{k=1,v_{kj}<0}^m\overline{z}%
_{ki}^cv_{kj}+\dsum\limits_{k=1,v_{kj}>0}^m\underline{z}_{ki}^cv_{kj}=%
\underline{r_{ij}}.
\]

Hence, we have proved that $p\widehat{z}_{ij}\in [\underline{r_{ij}},%
\overline{r_{ij}}]$ and we also have that $\underline{r_{ij}}$, $\overline{%
r_{ij}}$ are the projection of some vertex of the hypercube. Then we have
proved that the value of $\underline{r_{ij}}$ and $\overline{r_{ij}}$ are
given by the equation (7) and (8) respectively.

There are some very well known relations of duality between the eigenvectors
of $ZZ^t$ and $Z^tZ$, it is known that both matrix have the same $q$
eigenvalues strictly positives $\lambda _1,\lambda _2,\ldots ,\lambda _q$
and if we denote by $u_1,u_2,\ldots ,u_q$ the first $q$ eigenvectors of $Z^tZ
$, then the relations between the eigenvectors of $ZZ^t$ and $Z^tZ$ are show
in the equations (11) and (12):
\begin{equation}
u_\ell =\frac{Z^tv_\ell }{\sqrt{\lambda _\ell }}\text{ for }\ell =1,2,\ldots
,q.  \label{eq:eqd10}
\end{equation}

\begin{equation}
v_\ell =\frac{Zu_\ell }{\sqrt{\lambda _\ell }}\text{ for }\ell =1,2,\ldots
,q.  \label{eq:eqd11}
\end{equation}

With these ideas we propose three algorithms to apply a principal components
analysis that extend the one propose in \cite{ref3} in order to
produce a symbolic circle of correlation. We also propose an 3--th algorithm
to improve the time of the execution by considering which matrix is smaller $%
ZZ^t$ or $Z^tZ$.


\vspace{3mm}
\noindent {\large {\sc Algorithm 1: Principal Component Analysis with $ZZ^t$}}

\begin{description}
\item[Input]  :

\begin{itemize}
\item  $m=$number of symbolic objects.

\item  $n=$number of symbolic variables.

\item  The symbolic data table $X=\left(
\begin{array}{cccc}
\left[ \underline{x_{11}},\overline{x_{11}}\right]  & \left[ \underline{%
x_{12}},\overline{x_{12}}\right]  & \cdots  & \left[ \underline{x_{1n}},%
\overline{x_{1n}}\right]  \\
\left[ \underline{x_{21}},\overline{x_{21}}\right]  & \left[ \underline{%
x_{22}},\overline{x_{22}}\right]  & \cdots  & \left[ \underline{x_{2n}},%
\overline{x_{2n}}\right]  \\
\vdots  & \vdots  & \ddots  & \vdots  \\
\left[ \underline{x_{m1}},\overline{x_{m1}}\right]  & \left[ \underline{%
x_{m2}},\overline{x_{m2}}\right]  & \cdots  & \left[ \underline{x_{mn}},%
\overline{x_{mn}}\right]
\end{array}
\right) $.
\end{itemize}

\item[Output]  :

\begin{itemize}
\item  The symbolic correlation between the variables and the principal
components in the following matrix:

\[
R=\left(
\begin{array}{ccc}
\left[ \underline{R}(X^1,Y^1),\overline{R}(X^1,Y^1)\right]  & \cdots  &
\left[ \underline{R}(X^1,Y^n),\overline{R}(X^1,Y^n)\right]  \\
\vdots  & \ddots  & \vdots  \\
\left[ \underline{R}(X^n,Y^1),\overline{R}(X^n,Y^1)\right]  & \cdots  &
\left[ \underline{R}(X^n,Y^n),\overline{R}(X^n,Y^n)\right]
\end{array}
\right) .
\]

\item  The symbolic matrix with the first $q$ principal components:
\[
Y=\left(
\begin{array}{cccc}
\left[ \underline{y_{11}},\overline{y_{11}}\right]  & \left[ \underline{%
y_{12}},\overline{y_{12}}\right]  & \cdots  & \left[ \underline{y_{1q}},%
\overline{y_{1q}}\right]  \\
\left[ \underline{y_{21}},\overline{y_{21}}\right]  & \left[ \underline{%
y_{22}},\overline{y_{22}}\right]  & \cdots  & \left[ \underline{y_{2q}},%
\overline{y_{2q}}\right]  \\
\vdots  & \vdots  & \ddots  & \vdots  \\
\left[ \underline{y_{m1}},\overline{y_{m1}}\right]  & \left[ \underline{%
y_{m2}},\overline{y_{m2}}\right]  & \cdots  & \left[ \underline{y_{mq}},%
\overline{y_{mq}}\right]
\end{array}
\right) .
\]
\end{itemize}

\item[Step 1:]  Compute the matrix $X^c=(x_{ij}^c)_{\QATOPD. . {i=1,2,\ldots
,m}{j=1,2,\ldots ,n}}$ by:
\[
x_{ij}^c=\frac{\underline{x_{ij}}+\overline{x_{ij}}}2.
\]

\item[Step 2:]  Compute the matrix $Z=(z_{ij})_{\QATOPD. . {i=1,2,\ldots
,m}{j=1,2,\ldots ,n}}$ by:
\[
z_{ij}=\frac 1{\sqrt{m}}\frac{x_{ij}^c-\overline{X_j^c}}{\sigma _j^c}.
\]

\item[Step 3:]  Compute the matrix $\underline{Z}=(\underline{z}%
_{ij})_{\QATOPD. . {i=1,2,\ldots ,m}{j=1,2,\ldots ,n}}$ and $\overline{Z}=(%
\overline{z}_{ij})_{\QATOPD. . {i=1,2,\ldots ,m}{j=1,2,\ldots ,n}}$ by:
\[
\underline{z}_{ij}=\frac 1{\sqrt{m}}\frac{\underline{x_{ij}^{}}-\overline{%
X_j^c}}{\sigma _j^c}\text{,}
\]
\[
\overline{z}_{ij}^{}=\frac 1{\sqrt{m}}\frac{\overline{x_{ij}^{}}-\overline{%
X_j^c}}{\sigma _j^c}.
\]

\item[Step 4:]  Compute $H=ZZ^t$.

\item[Step 5:]  Compute the first $q$ eigenvectors $v_1,v_2,\ldots ,v_q$ of $%
H$ and the associated eigenvalues $\lambda _1\geq \lambda _2\geq \cdots \geq
\lambda _q>0$.

\item[Step 6:]  For $i=1,2,\ldots ,n$

\begin{description}
\item[Step 6.1:]  For $j=1,,2,\ldots ,q$ compute
\[
\underline{R}(X^i,Y^j)= \dsum\limits_{k=1,v_{kj}<0}^m\overline{z}%
_{ki}v_{kj}+\dsum\limits_{k=1,v_{kj}>0}^m\underline{z}_{ki}v_{kj} .
\]
\[
\overline{R}(X^i,Y^j)=\dsum\limits_{k=1,v_{kj}<0}^m\underline{z}%
_{ki}v_{kj}+\dsum\limits_{k=1,v_{kj}>0}^m\overline{z}_{ki}v_{kj} .
\]
\end{description}

\item[Step 7:]  For $i=1,2,\ldots ,n$

\begin{description}
\item[Step 7.1:]  For $j=1,2,\ldots ,q$ compute
\[
u_{ij}=\frac 1{\sqrt{\lambda _j}}\left(
\dsum\limits_{k=1}^mz_{ki}v_{kj}\right) .
\]
\end{description}

\item[Step 8:]  For $i=1,2,\ldots ,m$

\begin{description}
\item[Step 8.1:]  For $j=1,2,\ldots ,q$ compute
\[
\underline{y_{ij}}=\dsum\limits_{k=1,u_{kj<0}}^n\overline{z}%
_{ik}u_{kj}+\dsum\limits_{k=1,u_{kj>0}}^n\underline{z}_{ik}u_{kj}
\]
\[
\overline{y_{ij}}=\dsum\limits_{k=1,u_{kj<0}}^n\underline{z}%
_{ik}u_{kj}+\dsum\limits_{k=1,u_{kj>0}}^n\overline{z}_{ik}u_{kj}
\]
\end{description}

\item[Step 9:]  END of the algorithm.
\end{description}

The next algorithm extends the algorithm proposed in \cite{ref3}, it works
with the same variance--covariance matrix but we introduce some steps to compute the symbolic correlation
using duality relations in order to plot the symbolic circle of correlation.


\vspace{3mm}
\noindent {\large {\sc Algorithm 2: Principal Component Analysis Algorithm with $Z^tZ$}}

\begin{description}
\item[Input]  :

\begin{itemize}
\item  $m=$number of symbolic objects.

\item  $n=$number of symbolic variables.

\item  The symbolic data table $X=\left(
\begin{array}{cccc}
\left[ \underline{x_{11}},\overline{x_{11}}\right]  & \left[ \underline{%
x_{12}},\overline{x_{12}}\right]  & \cdots  & \left[ \underline{x_{1n}},%
\overline{x_{1n}}\right]  \\
\left[ \underline{x_{21}},\overline{x_{21}}\right]  & \left[ \underline{%
x_{22}},\overline{x_{22}}\right]  & \cdots  & \left[ \underline{x_{2n}},%
\overline{x_{2n}}\right]  \\
\vdots  & \vdots  & \ddots  & \vdots  \\
\left[ \underline{x_{m1}},\overline{x_{m1}}\right]  & \left[ \underline{%
x_{m2}},\overline{x_{m2}}\right]  & \cdots  & \left[ \underline{x_{mn}},%
\overline{x_{mn}}\right]
\end{array}
\right) $.
\end{itemize}

\item[Output]  :

\begin{itemize}
\item  The symbolic correlation between the variables and the principal
components in the following matrix:

\[
R=\left(
\begin{array}{ccc}
\left[ \underline{R}(X^1,Y^1),\overline{R}(X^1,Y^1)\right]  & \cdots  &
\left[ \underline{R}(X^1,Y^n),\overline{R}(X^1,Y^n)\right]  \\
\vdots  & \ddots  & \vdots  \\
\left[ \underline{R}(X^n,Y^1),\overline{R}(X^n,Y^1)\right]  & \cdots  &
\left[ \underline{R}(X^n,Y^n),\overline{R}(X^n,Y^n)\right]
\end{array}
\right) .
\]

\item  The symbolic matrix with the first $q$ principal components:
\[
Y=\left(
\begin{array}{cccc}
\left[ \underline{y_{11}},\overline{y_{11}}\right]  & \left[ \underline{%
y_{12}},\overline{y_{12}}\right]  & \cdots  & \left[ \underline{y_{1q}},%
\overline{y_{1q}}\right]  \\
\left[ \underline{y_{21}},\overline{y_{21}}\right]  & \left[ \underline{%
y_{22}},\overline{y_{22}}\right]  & \cdots  & \left[ \underline{y_{2q}},%
\overline{y_{2q}}\right]  \\
\vdots  & \vdots  & \ddots  & \vdots  \\
\left[ \underline{y_{m1}},\overline{y_{m1}}\right]  & \left[ \underline{%
y_{m2}},\overline{y_{m2}}\right]  & \cdots  & \left[ \underline{y_{mq}},%
\overline{y_{mq}}\right]
\end{array}
\right) .
\]
\end{itemize}

\item[Step 1:]  Compute the matrix $X^c=(x_{ij}^c)_{\QATOPD. . {i=1,2,\ldots
,m}{j=1,2,\ldots ,n}}$ by:
\[
x_{ij}^c=\frac{\underline{x_{ij}}+\overline{x_{ij}}}2.
\]

\item[Step 2:]  Compute the matrix $Z=(z_{ij})_{\QATOPD. . {i=1,2,\ldots
,m}{j=1,2,\ldots ,n}}$ by:
\[
z_{ij}=\frac 1{\sqrt{m}}\frac{x_{ij}^c-\overline{X_j^c}}{\sigma _j^c}.
\]

\item[Step 3:]  Compute the matrix $\underline{Z}=(\underline{z}%
_{ij})_{\QATOPD. . {i=1,2,\ldots ,m}{j=1,2,\ldots ,n}}$ and $\overline{Z}=(%
\overline{z}_{ij})_{\QATOPD. . {i=1,2,\ldots ,m}{j=1,2,\ldots ,n}}$ by:
\[
\underline{z}_{ij}=\frac 1{\sqrt{m}}\frac{\underline{x_{ij}^{}}-\overline{%
X_j^c}}{\sigma _j^c}\text{,}
\]
\[
\overline{z}_{ij}^{}=\frac 1{\sqrt{m}}\frac{\overline{x_{ij}^{}}-\overline{%
X_j^c}}{\sigma _j^c}.
\]

\item[Step 4:]  Compute $R=Z^tZ$.

\item[Step 5:]  Compute the first $q$ eigenvectors $u_1,u_2,\ldots ,u_q$ of $%
R$ and the associated eigenvalues $\lambda _1\geq \lambda _2\geq \cdots \geq
\lambda _q>0$.

\item[Step 6:]  For $i=1,2,\ldots ,m$

\begin{description}
\item[Step 6.1:]  For $j=1,2,\ldots ,q$ compute
\[
\underline{y_{ij}}=\dsum\limits_{k=1,u_{kj<0}}^n\overline{z}%
_{ik}u_{kj}+\dsum\limits_{k=1,u_{kj>0}}^n\underline{z}_{ik}u_{kj}
\]
\[
\overline{y_{ij}}=\dsum\limits_{k=1,u_{kj<0}}^n\underline{z}%
_{ik}u_{kj}+\dsum\limits_{k=1,u_{kj>0}}^n\overline{z}_{ik}u_{kj}
\]
\end{description}

\item[Step 7:]  For $i=1,2,\ldots ,m$

\begin{description}
\item[Step 7.1:]  For $j=1,2,\ldots ,q$ compute
\[
v_{ij}=\frac 1{\sqrt{\lambda _j}}\left(
\dsum\limits_{k=1}^mz_{ik}u_{kj}\right) .
\]
\end{description}

\item[Step 8:]  For $i=1,2,\ldots ,m$

\begin{description}
\item[Step 8.1:]  For $j=1,2,\ldots ,q$ compute
\[
\underline{R}(X^i,Y^j)=\dsum\limits_{k=1,v_{kj}<0}^m\overline{z}%
_{ki}v_{kj}+\dsum\limits_{k=1,v_{kj}>0}^m\underline{z}_{ki}v_{kj} .
\]
\[
\overline{R}(X^i,Y^j)=\dsum\limits_{k=1,v_{kj}<0}^m\underline{z}%
_{ki}v_{kj}+\dsum\limits_{k=1,v_{kj}>0}^m\overline{z}_{ki}v_{kj} .
\]
\end{description}

\item[Step 9:]  END of the algorithm.
\end{description}

The size of the matrix $ZZ^t$ is $m\times m$ while the size of $Z^tZ$ is $%
n\times n$, sometimes $ZZ^t$ is very big and $Z^tZ$ is very small, in this
case is better to use the algorithm 2 than the algorithm 1, or inversely $%
Z^tZ$ is very big and $ZZ^t$ is very small then it is faster the algorithm 1
than the algorithm 2. Hence, considering if $m\leq n$ or not we propose the
algorithm 3.


\vspace{3mm}
\noindent {\large {\sc Algorithm 3: Principal Component Analysis Optimal Algorithm}}

\begin{description}
\item[Input]  :

\begin{itemize}
\item  $m=$number of symbolic objects.

\item  $n=$number of symbolic variables.

\item  The symbolic data table $X=\left(
\begin{array}{cccc}
\left[ \underline{x_{11}},\overline{x_{11}}\right]  & \left[ \underline{%
x_{12}},\overline{x_{12}}\right]  & \cdots  & \left[ \underline{x_{1n}},%
\overline{x_{1n}}\right]  \\
\left[ \underline{x_{21}},\overline{x_{21}}\right]  & \left[ \underline{%
x_{22}},\overline{x_{22}}\right]  & \cdots  & \left[ \underline{x_{2n}},%
\overline{x_{2n}}\right]  \\
\vdots  & \vdots  & \ddots  & \vdots  \\
\left[ \underline{x_{m1}},\overline{x_{m1}}\right]  & \left[ \underline{%
x_{m2}},\overline{x_{m2}}\right]  & \cdots  & \left[ \underline{x_{mn}},%
\overline{x_{mn}}\right]
\end{array}
\right) $.
\end{itemize}

\item[Output]  :

\begin{itemize}
\item  The symbolic correlation between the variables and the principal
components in the following matrix:

\[
R=\left(
\begin{array}{ccc}
\left[ \underline{R}(X^1,Y^1),\overline{R}(X^1,Y^1)\right]  & \cdots  &
\left[ \underline{R}(X^1,Y^n),\overline{R}(X^1,Y^n)\right]  \\
\vdots  & \ddots  & \vdots  \\
\left[ \underline{R}(X^n,Y^1),\overline{R}(X^n,Y^1)\right]  & \cdots  &
\left[ \underline{R}(X^n,Y^n),\overline{R}(X^n,Y^n)\right]
\end{array}
\right) .
\]

\item  The symbolic matrix with the first $q$ principal components:
\[
Y=\left(
\begin{array}{cccc}
\left[ \underline{y_{11}},\overline{y_{11}}\right]  & \left[ \underline{%
y_{12}},\overline{y_{12}}\right]  & \cdots  & \left[ \underline{y_{1q}},%
\overline{y_{1q}}\right]  \\
\left[ \underline{y_{21}},\overline{y_{21}}\right]  & \left[ \underline{%
y_{22}},\overline{y_{22}}\right]  & \cdots  & \left[ \underline{y_{2q}},%
\overline{y_{2q}}\right]  \\
\vdots  & \vdots  & \ddots  & \vdots  \\
\left[ \underline{y_{m1}},\overline{y_{m1}}\right]  & \left[ \underline{%
y_{m2}},\overline{y_{m2}}\right]  & \cdots  & \left[ \underline{y_{mq}},%
\overline{y_{mq}}\right]
\end{array}
\right) .
\]
\end{itemize}

\item[Step 1:]  If $m\leq n$ then we apply algorithm 2 else we apply
algorithm 2.

\item[Step 2:]  END of the algorithm.
\end{description}

\section{Experimental evaluation}

\subsection{Ichino interval data set}

To illustrate the symbolic circle of correlations we use Ichinos'data (oils
and fats) that we present in the Table \ref{tabla1}. Each row of the data
table refers to a class of oil described by 4 quantitative interval type
variables, ``Specific gravity'', ``Freezing point'', ``Iodine value'' and
``Saponification''.

\begin{table}[tb]
\begin{center}
\begin{tabular}{||l||c|c|c|c||}
\hline
& GRA & FRE & IOD & SAP \\ \hline\hline
Linsed (L) & $\left[ 0.93,0.935\right] $ & $\left[ -27,-18\right] $ & $%
\left[ 170,204\right] $ & $\left[ 118,196\right] $ \\ \hline
Perilla (P) & $\left[ 0.93,0.937\right] $ & $\left[ -5,-4\right] $ & $\left[
192,208\right] $ & $\left[ 188,197\right] $ \\ \hline
Cotton (Co) & $\left[ 0.916,0.918\right] $ & $\left[ -6,-1\right] $ & $%
\left[ 99,113\right] $ & $\left[ 189,198\right] $ \\ \hline
Sesame (S) & $\left[ 0.92,0.926\right] $ & $\left[ -6,-4\right] $ & $\left[
104,116\right] $ & $\left[ 187,193\right] $ \\ \hline
Camellia (Ca) & $\left[ 0.916,0.917\right] $ & $\left[ -25,-15\right] $ & $%
\left[ 80,82\right] $ & $\left[ 189,193\right] $ \\ \hline
Olive (O) & $\left[ 0.914,0.919\right] $ & $\left[ 0,6\right] $ & $\left[
79,90\right] $ & $\left[ 187,196\right] $ \\ \hline
Beef (B) & $\left[ 0.86,0.87\right] $ & $\left[ 30,38\right] $ & $\left[
40,48\right] $ & $\left[ 190,199\right] $ \\ \hline
Hog (H) & $\left[ 0.858,0.864\right] $ & $\left[ 22,32\right] $ & $\left[
53,77\right] $ & $\left[ 190,202\right] $ \\ \hline
\end{tabular}
\end{center}
\caption{Oils and Fats data table.}
\label{tabla1}
\end{table}

The symbolic correlations that we got using the Algorithm 3 are presented in
Table \ref{tablad1} and the classical correlations between the gravity
center of variables and the gravity center of the principal components (for
the centers method) are presented in the Table \ref{tablad2}. Notice you
that with this method it is fulfilled that the classic correlations are
always contained in the interval that represents the symbolic correlation.

\begin{table}[tb]
\begin{center}
\begin{tabular}{||c||c|c|c|c||}
\hline
& PC1 & PC2 & PC3 & PC4 \\ \hline\hline
GRA & $[0.827,1.000]$ & $[-0.443,-0.265]$ & $[-0.038, 0.087]$ & $%
[-0.238,-0.084]$ \\ \hline
FRE & $[-1.000,-0.760]$ & $[0.044, 0.372]$ & $[-0.428,-0.220]$ & $[-0.288,
0.019]$ \\ \hline
IOD & $[0.726, 1.000]$ & $[-0.124, 0.191]$ & $[-0.565,-0.401]$ & $[-0.024,
0.161]$ \\ \hline
SAP & $[-1.000, 0.190]$ & $[-1.000, 0.371]$ & $[-0.442, 0.163]$ & $[-0.231,
0.325]$ \\ \hline
\end{tabular}
\end{center}
\caption{Symbolic correlations between the variables and principal
components with Duality Center Method.}
\label{tablad1}
\end{table}

\begin{table}[tb]
\begin{center}
\begin{tabular}{||c||c|c|c|c||}
\hline
& PC1 & PC2 & PC3 & PC4 \\ \hline\hline
GRA & $0.9210665$ & $-0.3537703$ & $0.0246894$ & $-0.1608524$ \\ \hline
FRE & $-0.9130654$ & $0.2080771$ & $-0.3238118$ & $-0.1347643$ \\ \hline
IOD & $0.8724116$ & $0.0337627$ & $-0.4827661$ & $0.0685206$ \\ \hline
SAP & $-0.7354523$ & $-0.6613331$ & $-0.1397354$ & $0.0471425$ \\ \hline
\end{tabular}
\end{center}
\caption{Classical correlations between the variables and principal
components with Duality Center Method.}
\label{tablad2}
\end{table}

\begin{figure}[tbp]
\begin{center}
\includegraphics[height=8cm]{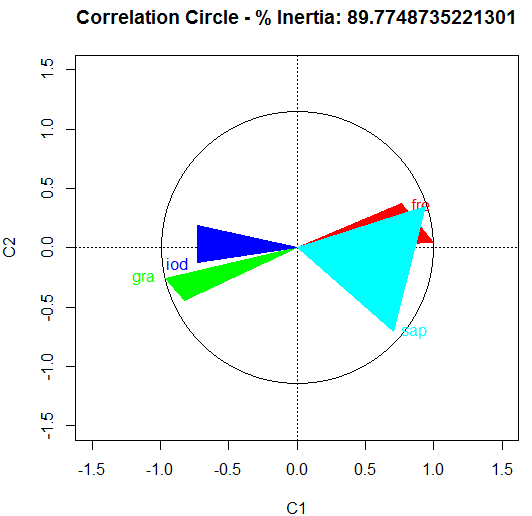}
\end{center}
\caption{Minimum and maximum distances between 2 hyper--rectangles.}
\label{figcd1}
\end{figure}

\begin{figure}[tbp]
\begin{center}
\includegraphics[height=8cm]{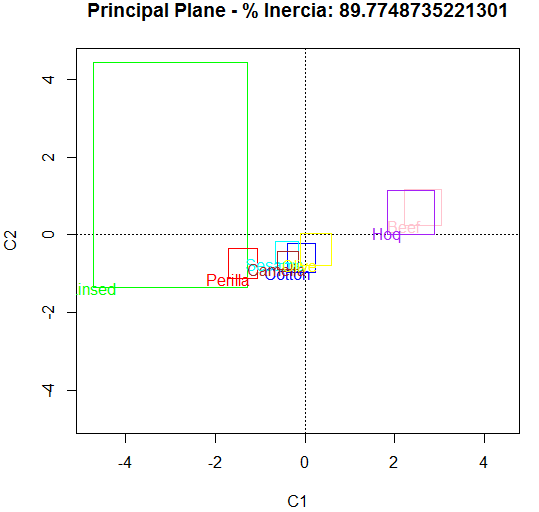}
\end{center}
\caption{Symbolic principal plane with Duality Centers Method.}
\label{figpd1}
\end{figure}

The symbolic circle of correlation to the oils and fats data got by the
Duality Center Method is shown in Figure \ref{figcd1}. The principal plane
got by the Duality Center Method and corresponding to this correlation
circle is presented in Figure \ref{figpd1} and principal components are
presented in the Table \ref{tablad2_1}.

\begin{table}[tb]
\begin{center}
\begin{tabular}{||l||c|c|c|c||}
\hline
& PC1 & PC2 & PC3 & PC4 \\ \hline\hline
L & $[ 1.275, 4.733]$ & $[-1.353, 4.428]$ & $[-1.025, 1.289]$ & $[-0.989,
0.989]$ \\ \hline
P & $[ 1.059, 1.701]$ & $[-1.128,-0.343]$ & $[-1.508,-1.046]$ & $[-0.134,
0.334]$ \\ \hline
Co & $[-0.236, 0.399]$ & $[-0.969,-0.213]$ & $[-0.170, 0.368]$ & $[-0.246,
0.204]$ \\ \hline
S & $[ 0.154, 0.658]$ & $[-0.745,-0.179]$ & $[-0.027, 0.342]$ & $[-0.369,
0.028]$ \\ \hline
Ca & $[ 0.151, 0.613]$ & $[-0.881,-0.437]$ & $[ 0.807, 1.204]$ & $[ 0.113,
0.538]$ \\ \hline
O & $[-0.594, 0.100]$ & $[-0.775, 0.043]$ & $[ 0.019, 0.545]$ & $%
[-0.645,-0.101]$ \\ \hline
B & $[-3.046,-2.226]$ & $[ 0.234, 1.162]$ & $[-0.392, 0.152]$ & $[-0.530,
0.193]$ \\ \hline
H & $[-2.900,-1.841]$ & $[ 0.020, 1.135]$ & $[-0.729, 0.171]$ & $[-0.105,
0.720]$ \\ \hline
\end{tabular}
\end{center}
\caption{Principal components with Duality Center Method.}
\label{tablad2_1}
\end{table}

\subsection{US Murder interval data set}

These data were taken from UCI Machine Learning Repository, consult \cite{Bache2013} and  \cite{Rod2014}. The idea is to study this murders table in $n = 1994$ communities of the United State, for this we have $p = 103$ variables measured over each one of these communities. The objective is to predict the mortality rate (ViolentCrimesPerPop) as a linear function of these $p = 103$ predictors.

The first action taken to analyze this data set was to convert it in a symbolic data set where all predictors are of the interval type, taking for this as concept the variable {\tt State}, that is, each state of US will be a statistic unit that we will study. We will thus have 46 rows (states present in US Murder data set) and 102 interval-valued predictors. Table \ref{Tab8} presents partially the US Murder classic data set, this table being $1994 \times 103$ and Table \ref{Tab9} presents partially the US Murder interval data set, the size of this table is $46 \times 102$. Chapter 5 of the book \cite{BockDiday2000}  explains in detail how to transform a classic data table in a symbolic data table. 

\begin{table}[ht] 
\begin{footnotesize}
\begin{center} 
\begin{tabular}{|c|c|c|c|c|c|c|c|c|}
  \hline
\rowcolor[rgb]{0,1,1}  N & state & fold & population & householdsize & racepctblack & racePctWhite & racePctAsian & $\cdots$ \\    \hline  
\rowcolor[gray]{0.9}  1 &  8  &  1 &      0.19 &         0.33   &      0.02    &     0.90     &    0.12 & $\cdots$  \\   \hline
\rowcolor[gray]{0.9}  2  &  53 &   1   &    0.00   &       0.16  &       0.12  &       0.74    &     0.45 & $\cdots$  \\   \hline
\rowcolor[gray]{0.9}  3   & 24  &  1   &    0.00    &      0.42  &       0.49    &     0.56     &    0.17 & $\cdots$  \\   \hline
\rowcolor[gray]{0.9}  4  &  34   & 1   &    0.04    &      0.77  &       1.00    &     0.08      &   0.12 & $\cdots$  \\   \hline
\rowcolor[gray]{0.9}  5   & 42   & 1   &    0.01     &     0.55   &      0.02    &     0.95     &    0.09 & $\cdots$  \\   \hline
\rowcolor[gray]{0.9}  6   &  6  &  1   &    0.02      &    0.28    &     0.06     &    0.54      &   1.00 & $\cdots$  \\   \hline
\rowcolor[gray]{0.9}  7   & 44   & 1  &     0.01      &    0.39    &     0.00    &     0.98      &   0.06 & $\cdots$  \\   \hline
\rowcolor[gray]{0.9}  8   &  6   & 1   &    0.01      &    0.74     &    0.03     &    0.46     &    0.20 & $\cdots$  \\   \hline
\rowcolor[gray]{0.9}  $\vdots$ &  $\vdots$ & $\vdots$ & $\vdots$ & $\vdots$ & $\vdots$ & $\vdots$ & $\vdots$ & $\cdots$  \\  \hline
\rowcolor[gray]{0.9}  1994   &  6  & 10   &    0.2      &    0.78     &    0.14    &    0.46     &    0.24 & $\ddots$  \\   \hline
\end{tabular}
\end{center}
\caption{US Murder classic data set.}
\label{Tab8}
\end{footnotesize}
\end{table}

\begin{table}[ht] 
\begin{small}
\begin{center} 
\begin{tabular}{|c|c|c|c|c|c|c|c|}
  \hline
\rowcolor[rgb]{0,1,1}   state & fold & population & householdsize & racepctblack & racePctWhite & racePctAsian & $\cdots$ \\    \hline  
\rowcolor[gray]{0.9}  1 &  [1,10]   &  [0,0.41] & [0.23,0.67] & [0,1.00] &  [0,0.99]  & [0,0.21] &  $\cdots$  \\   \hline
\rowcolor[gray]{0.9}  2 &  [4,9]   &  [0.03,0.35] & [0.46,0.53] & [0.02,0.25] &  [0.58,0.71]  & [0.2,0.3] &  $\cdots$  \\   \hline
\rowcolor[gray]{0.9}  3 &  [1,10]   &  [0,1] & [0.23,0.98] & [0,0.23] &  [0.37,0.97]  & [0.02,0.32] &  $\cdots$  \\   \hline
\rowcolor[gray]{0.9}  4 &  [1,10]   &  [0,0.27] & [0.22,0.59] & [0,1] &  [0.12,1]  & [0.01,0.25] &  $\cdots$  \\   \hline
\rowcolor[gray]{0.9}  5 &  [1,10]   &  [0,1] & [0,1] & [0,1] &  [0,0.96]  & [0.03,1] &  $\cdots$  \\   \hline
\rowcolor[gray]{0.9}  6 &  [1,9]   &  [0,0.74] & [0.21,0.6] & [0,0.25] &  [0.58,0.95]  & [0.01,0.24] &  $\cdots$  \\   \hline
\rowcolor[gray]{0.9}  $\vdots$ &  $\vdots$ & $\vdots$ & $\vdots$ & $\vdots$ & $\vdots$ & $\vdots$  & $\cdots$  \\  \hline
\rowcolor[gray]{0.9}  46 &  [3,9]   &  [0,0.06] & [0.29,0.67] & [0,0.06] &  [0.85,0.97]  & [0.02,0.14] &  $\ddots$  \\   \hline
\end{tabular}
\end{center}
\caption{US Murder interval data set.}
\label{Tab9}
\end{small}
\end{table}

\section{The duality problem in Tops Method}

\markright{The duality problem in Tops Method}

It is impossible to extend the Tops Method algorithm proposed in \cite{ref3}
using duality
relation to compute the circle of correlation, because we can project the
center of gravity of the variables but we can not project the hypercube
variables. To project the center of gravity of the variables we need to
calculate the eigenvectors of $ZZ^t$ using the relation $v_\ell =\frac{%
Zu_\ell }{\sqrt{\lambda _\ell }}$, this is possible because the size of $Z$
is $m\cdot 2^n\times n$ and the size of $u_\ell $ is $n\times n$ then the
size of the matrix $V=[v_1|v_2|\cdots |v_q]$ is $m\cdot 2^n\times q$, so it
is possible to get the coordinates of the variables computing $Z^tV$. But,
to compute the symbolic projection of the variables like rectangles we
should to compute $\underline{R}(X^i,Y^j)=\dsum\limits_{k=1,v_{kj}<0}^m%
\overline{z}_{ki}v_{kj}+\dsum\limits_{k=1,v_{kj}>0}^m\underline{z}%
_{ki}v_{kj} $ and $\overline{R}(X^i,Y^j)=\dsum\limits_{k=1,v_{kj}<0}^m%
\underline{z}_{ki}v_{kj}+\dsum\limits_{k=1,v_{kj}>0}^m\overline{z}_{ki}v_{kj}
$ and it is not possible because the size of $\overline{z}_i=(\overline{z}%
_{11},\overline{z}_{21},\ldots ,\overline{z}_{m1})$ and $v_j=(v_{1j},v_{2j},%
\ldots ,v_{m\cdot 2^nj})$ are clearly different.

\end{document}